\documentclass[10pt,conference]{IEEEtran}
\usepackage{subfig}
\usepackage{afterpage}
\usepackage{graphicx}
\usepackage{makecell}
\usepackage{xcolor}
\usepackage{dingbat}
\usepackage{amssymb}
\usepackage{listings}
\usepackage{pifont}

\definecolor{mGreen}{rgb}{0,0.6,0}
\definecolor{mGray}{rgb}{0.5,0.5,0.5}
\definecolor{mPurple}{rgb}{0.58,0,0.82}
\definecolor{backgroundColour}{rgb}{0.95,0.95,0.95}

\lstdefinestyle{CStyle}{
    backgroundcolor=\color{backgroundColour},   
    commentstyle=\color{mGreen},
    keywordstyle=\color{blue},
    numberstyle=\tiny\color{magenta},
    stringstyle=\color{mPurple},
    basicstyle=\footnotesize,
    breakatwhitespace=false,         
    breaklines=true,                 
    captionpos=b,                    
    keepspaces=true,                 
    numbers=left,                    
    numbersep=5pt,                  
    showspaces=false,                
    showstringspaces=false,
    showtabs=false,                  
    tabsize=2,
    language=C
}

\usepackage{xcolor}
\newcommand{\greencheck}{{\color{green}\checkmark}}
\newcommand{\orangecheck}{{\color{orange}\checkmark}}
\newcommand{\redcheck}{{\color{red}\xmark}}

\newcommand{\xmark}{\ding{55}}%

\begin{document}

\title{STONNE: A Detailed Architectural Simulator for Flexible Neural Network Accelerators}

\author{\IEEEauthorblockN{Francisco Mu\~noz-Mart\'inez, Manuel E. Acacio}
\IEEEauthorblockA{Universidad de Murcia\\Murcia, SPAIN
\\\texttt{\{francisco.munoz2,meacacio\}@um.es}}
\and
\IEEEauthorblockN{Jos\'e L. Abell\'an}
\IEEEauthorblockA{Universidad Cat\'olica San Antonio\\ Murcia, SPAIN\\
\texttt{jlabellan@ucam.edu}}
\and
\IEEEauthorblockN{Tushar Krishna}
\IEEEauthorblockA{Georgia Institute of Technology \\
Atlanta, Georgia, USA\\
\texttt{tushar@ece.gatech.edu
}}}



\maketitle
\begin{abstract}
The design of specialized architectures for accelerating the inference procedure of Deep Neural Networks (DNNs) is a booming area of research nowadays. First-generation rigid proposals have been rapidly replaced by more advanced flexible accelerator architectures able to efficiently support a variety of layer types and dimensions. As the complexity of the designs grows, it is more and more appealing for researchers to have cycle-accurate simulation tools at their disposal to allow for fast and accurate design-space exploration, and rapid quantification of the efficacy of architectural enhancements during the early stages of a design. To this end, we present STONNE (\underline{S}imulation \underline{TO}ol of \underline{N}eural \underline{N}etwork \underline{E}ngines), a cycle-accurate, highly-modular and  highly-extensible simulation framework that enables end-to-end evaluation of flexible accelerator architectures running complete contemporary DNN models. We use STONNE to model the recently proposed MAERI architecture and show how it can closely approach the performance results of the publicly available BSV-coded MAERI implementation. Then, we conduct a comprehensive evaluation and demonstrate that the folding strategy implemented for MAERI results in very low compute unit utilizations (25\% on average across 5 DNN models) which in the end translates into poor performance.  
\end{abstract}


\section{Introduction}
\label{section:introduction}


Contemporary Deep Neural Networks (DNNs) are organized as a large number of layers (e.g., convolution and fully-connected), each composed of a large set of neurons. Depending on the type of layer, each neuron performs a simple weighted addition of (or some of) the values obtained in the preceding layer. During the inference phase the already trained DNN model is used to make a prediction.

The difficulty in processing these workloads does not stem from the type of operations to be performed (simple Multiply-Accumulate operations or MACs with 8-bit operands might suffice~\cite{Survey2017}) but from the vast amount of MAC operations involved in the inference procedure of DNNs (e.g., 3.9 billions in ResNet-50). As a result, typical DNN layers are excessively large to be executed by an edge-computing accelerator in a single step. 
So, when processing a single layer, their neurons are grouped in smaller tiles that define the pattern in which the neurons' inputs, weights, and intermediate outputs (partial sums or psums) are delivered and reused within the accelerator's functional units. This pattern is called dataflow and determines the energy efficiency of an accelerator architecture when processing a certain DNN layer~\cite{Survey2017}.
First-generation DNN inference accelerators focused their designs on fixed-size clusters of multipliers-and-accumulate units interconnected by means of a fixed on-chip network fabric specifically tailored to efficiently support a particular dataflow. Unfortunately, the inability of these designs to adapt well to the varying morphology among contemporary DNNs, and more importantly, among different layers within the same DNN (varying layer dimensions and types~\cite{Eyeriss2017}), limits their potential advantages (low compute unit utilization and low reuse of data that is translated into low energy efficiency). To overcome this limitation, recent proposals such as FlexFlow~\cite{FlexFlow2017}, MAERI~\cite{MAERI2018} and SIGMA~\cite{SIGMA2020} advocate using flexible DNN accelerator fabrics, which can be reconfigured to efficiently map different dataflows and partitions through the creation of dynamic-size clusters in the same hardware substrate. Of course, this flexibility comes at the cost of increased accelerator complexity that urges for a more exhaustive design-space exploration for fine tuning before building the particular ASIC-based or FPGA-based DNN accelerator prototype.


Traditionally, architectural simulators have become an integral part of the computer architecture research and design process, since they permit fast and accurate design-space exploration and rapid quantification of the efficacy of architectural enhancements in the early stages of a design, and have been extensively used during the design process of CPU and GPU architectures~\cite{gem5,mgpusim}. 
However, and quite surprisingly, the same has not taken place until now for inference accelerator architectures. To the best of our knowledge, there is still no detailed open-source architectural simulator for extensive and accurate design-space exploration of next-generation flexible DNN inference accelerators. In this work we present STONNE (\underline{S}imulation \underline{TO}ol of \underline{N}eural \underline{N}etwork \underline{E}ngines), a cycle-accurate, highly-modular and highly-extensible simulator aimed to bridge this gap. 

\begin{table}[t!]
\caption{State-of-the-art simulators for DNN architectures.}
\vspace{-0.25cm}
\label{table:state_of_the_art}
\begin{scriptsize}
\begin{center}
{
\begin{tabular}{|c|c|c|c|}\hline
                    & \makecell{\textbf{End-to-End}\\\textbf{evaluation}}  &  \makecell{\textbf{Easy} \textbf{to}\\\textbf{Extend}}    &   \makecell{\textbf{Flexible}\\\textbf{architecture}}   \\\hline
\makecell{MAGNet~\cite{MAGNet2019}\\DNNBuilder~\cite{DNNBuilder2019}} & \redcheck   &    \redcheck      &   \redcheck \\\hline
\makecell{MAERI BSV~\cite{MaeriCode}}                     & \redcheck  & \redcheck   & \orangecheck \\\hline
\makecell{TVM~\cite{TVM2018}}                     & \greencheck  & \redcheck   & \redcheck \\\hline
\makecell{SCALE-Sim~\cite{Samajdar2019}\\MAESTRO~\cite{MAESTRO2019}}      & \redcheck   &   \greencheck    &   \redcheck \\\hline
\makecell{SMAUG~\cite{SMAUG2019}}         & \greencheck & \greencheck & \redcheck  \\\hline
\makecell{STONNE}                         & \greencheck & \greencheck & \greencheck \\\hline
\end{tabular}
}
\vspace{-0.55cm}
\end{center}
\end{scriptsize}
\end{table}

Table~\ref{table:state_of_the_art} shows a qualitative comparison of STONNE with respect to contemporary publicly available simulators for DNN inference accelerators. 
As we can see, unlike STONNE, existing simulators were originally developed for first-generation DNN accelerators and do not give support for simulating flexible DNN architectures. This is at least in part because it is not possible, without significant ``heavy lifting'' to extend these simulation tools to support next-generation DNN accelerators, since they were tailored to specifically simulate a certain type of rigid architecture (e.g., a systolic array as in Google TPU~\cite{jouppi-2017}). 
Among all the alternatives, only the MAERI BSV model and SMAUG claim to model early flexible architectures. However, none of them really allows neither for efficient design-space exploration nor for rapid quantification of the impact that modifications in the architecture of a flexible inference accelerator would have on both performance and energy consumption. 

To motivate how STONNE can be used for research in the design and implementation of flexible next-generation DNN accelerators, we consider the case of the state-of-the-art MAERI architecture (further details in Section~\ref{section:architecture}). In particular, we examine how performance is affected by both the number of multipliers in the MAERI architecture (64 and 128 multipliers) and the strategy currently being used to handle folding\footnote{Folding is utilized when a neuron needs more multiplication operations than the number of multiplier units available in hardware. Then, the neuron is ``folded'' to be processed in several sequential steps and partial results should be accumulated and taken at inter-steps boundaries.}. The latter is done through the execution of five complete DNN models: AlexNet~\cite{Krizhevsky2012}, MobileNets~\cite{MobileNets2017}, Squeezenet~\cite{Squeezenet2016}, Resnet-50~\cite{ResNet2012} and VGG-16~\cite{VGG162015}. To configure the MAERI model, we use the best tile configuration for every layer of every DNN by using mRNA~\cite{mRNA2019}, the search exploration tool to configure the MAERI architecture. Additionally, we assume perfect bandwidth (no contention) between memory and the MAERI's processing elements (a fabric of multipliers with a tree-based reduction tree of adders to efficiently map MAC operations onto the hardware substrate).  


\begin{figure}[t!]
        \begin{center}
                \includegraphics[width=0.5\textwidth]{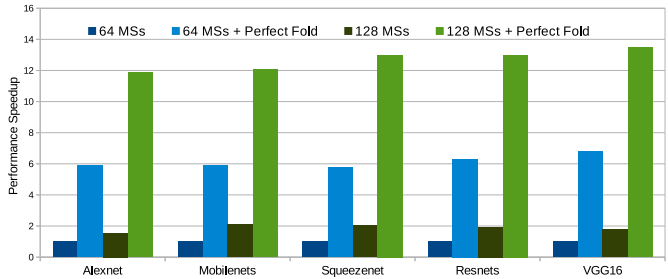}
        \end{center}
        \vspace{-0.2cm}
        \caption{Performance results for simple design-space exploration in a simulated MAERI accelerator: speedup obtained by doubling compute resources (128 multipliers) and speedup that would be obtained for an ideal implementation of folding (\texttt{+ Perfect Fold}).}
        \label{fig:performance_speedup}
\end{figure}

As we can see in Figure~\ref{fig:performance_speedup}, when the number of multipliers in MAERI doubles (\texttt{128 MSs}), the performance improvement almost achieves ideal scaling (an average of 1.88$\times$). This significant speedup is obtained by doubling the amount of hardware resources for computing MAC operations, which might not be a feasible design decision to increase performance in highly-constrained low-end edge-computing devices. A much more cost-effective solution to increase performance in MAERI is optimizing the implementation of folding. Ideal folding with both 64 and 128 multipliers (\texttt{64/128 MSs + Perfect Fold}) reports impressive performance benefits. In particular, average speedups of 6.1$\times$ and 12.6$\times$ with respect to the baseline (\texttt{64 MSs}). In Section~\ref{section:validation} we will dig into the details and explain the reasons behind this performance bottleneck.

We see the following contributions in this work:
\begin{itemize}
    \item We present for the first time STONNE, a cycle-accurate architectural simulator for flexible DNN inference accelerators that features high modularity, high configurability and end-to-end evaluation.
    \item We model, configure, implement and validate a MAERI architecture in STONNE by using end-to-end evaluation with state-of-the-art DNN models. We validate our implementation against a publicly available real BlueSpec SystemVerilog (BSV)-based prototype~\cite{MaeriCode}. We obtain an average difference of just 15\% in terms of total executed cycles and we identify where this difference comes from.
    \item We conduct a comprehensive characterization to illustrate the key benefits of STONNE. In particular, we demonstrate the significant performance bottleneck of the BSV-based prototype of MAERI when implementing folding.  By using the statistics reported by STONNE, we identify that the main root of such a performance bottleneck lies in the current design of the MAERI's reduction network (RN).
\end{itemize}

The rest of the work is organized as follows. First, Section~\ref{section:design} explains the organization of STONNE. After that, Section~\ref{section:architecture} describes the family of flexible accelerator architectures that STONNE simulates. Section~\ref{section:validation} demonstrates the accuracy and potential of the tool and evaluates a performance bottleneck found in a typical flexible DNN architecture. Finally, Section~\ref{section:conclusions} presents the main conclusions of this work and outlines the ongoing work.  

\section{STONNE FRAMEWORK} 
\label{section:design}
\begin{figure*}[t!]
        \begin{center}
                \includegraphics[width=0.8\textwidth]{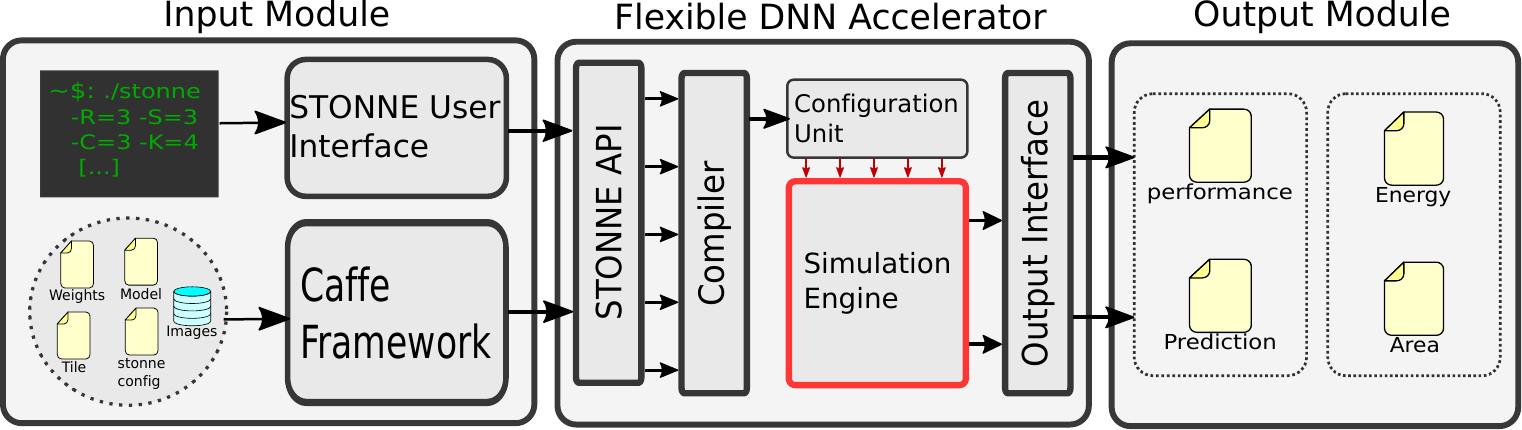}
        \end{center}
        \vspace{-0.2cm}
        \caption{High-level diagram of the {STONNE} framework.}
        \label{fig:simulator_scheme}
\end{figure*}

 {STONNE} is a highly-configurable cycle-accurate next-generation DNN accelerator simulator. The simulator has been developed in C++ and allows for end-to-end evaluation as it is connected with Caffe DL framework. Current version of the simulator can fully execute any relevant DNN model  and, as the GRASP and SOLID programming principles of object-oriented design have been followed to build the simulator, STONNE is highly extensible and can be easily modified to support any particularity of any DNN model (e.g., different type of layers). In addition, it can be easily extended to model different architectures of flexible DNN accelerators, tile configuration mappings and dataflows.
 
 \subsection{STONNE Organization}
 
 Figure~\ref{fig:simulator_scheme} illustrates a high-level diagram of {STONNE} with the three major components involved in the end-to-end simulation flow. First, the \textbf{Input Module} determines the layer to be run, creates an instance of the simulator and loads the parameters of the layer and the initial inputs and weights onto the architecture. Once the architecture has been configured, the \textbf{Flexible DNN Accelerator} module carries out the detailed cycle-by-cycle simulation of the layer, collecting statistics during the process. Once the simulator finishes, 
 the \textbf{Output Module} takes in the values of the counters collected by the simulator and produces different files with the statistics of the execution. 
 
 Next, we further describe the details of each module:   


\textbf{(1) Flexible DNN Accelerator}: This constitutes the principal block of {STONNE} (see the central block in Figure~\ref{fig:simulator_scheme}), and it is mainly composed of the modeled architecture of the flexible DNN accelerator (\texttt{Simulation Engine}) to simulate, whose details are further explained in Section~\ref{section:architecture}. The accelerator is interfaced by means of the \texttt{STONNE API} that allows users to create an instance of the simulator according to a hardware configuration file, load the layer and tile parameters, and load the weights and inputs onto the memory of the simulator.
Once all these parameters have been defined, the \texttt{Compiler} generates all the control signals that configure the architecture through the \texttt{Configuration Unit}. Then, the simulator starts the execution and the results and statistics being collected are reported through an output interface.

\textbf{(2) Input Module}: Due to the flexibility that the \texttt{STONNE API} provides, the simulator can be fed easily using any of the well-known DL frameworks already available. In this work, we have modified the Caffe DL framework (see left block in Figure~\ref{fig:simulator_scheme}) to connect it to the simulator so that it is able to run an instance of the \texttt{Simulation Engine} (e.g., MAERI) transparently to the user. This way, a Caffe user just needs to select the typical \textit{.caffemodel} file with the weights, choose the inputs\footnote{Throughout this work, we use STONNE to characterize the inference process of several contemporary DNN models aimed to image classification.} (e.g., a set of images) and briefly modify each layer block defined in  the \textit{.prototxt} DNN model file to specify the layers to be simulated, the path of the hardware configuration file with the parameters of the architecture to simulate (e.g., the number of multipliers) and the tile configuration for every layer. After Caffe is launched with those defined parameters, it takes the control and creates an instance of STONNE. Then, Caffe drives a layer-by-layer execution, sending the configuration parameters for every layer, copying the weights and the inputs of that layer onto the simulator memory, and copying back the results after the simulator finishes and produces the statistics file. This process is repeated for every layer until the end of the execution, producing the final prediction for each input (thus performing the whole inference process).

Furthermore, since Caffe requires a more complicated installation and use, apart from this mode of execution, we have also enabled the \texttt{STONNE User Interface} that facilitates the execution of STONNE. Through this mode, the user is presented with a prompt to load any layer and tile parameters onto a selected instance of the  simulator and run it with random weights and input values. This mode allows for faster executions and hence facilitates faster prototyping and debugging.

\textbf{(3) Output module}: Once a simulation for a certain layer has been completed, this module is used for reporting simulation statistics such as performance, compute unit utilization, number of accesses to SRAM, wires and FIFOs, etc. Besides, this output module also reports the amount of energy consumed and the on-chip area required by the simulated architecture. Currently, we are extending the simulator to provide such area and energy numbers. Moreover, since the STONNE simulator is a back-end compute platform of Caffe, it also outputs the result of the prediction when running a certain DNN model for certain input data.

\subsection{Flexible DNN Accelerator Architecture}

\begin{figure}[t!]
        \begin{center}
                \includegraphics[width=0.4\textwidth]{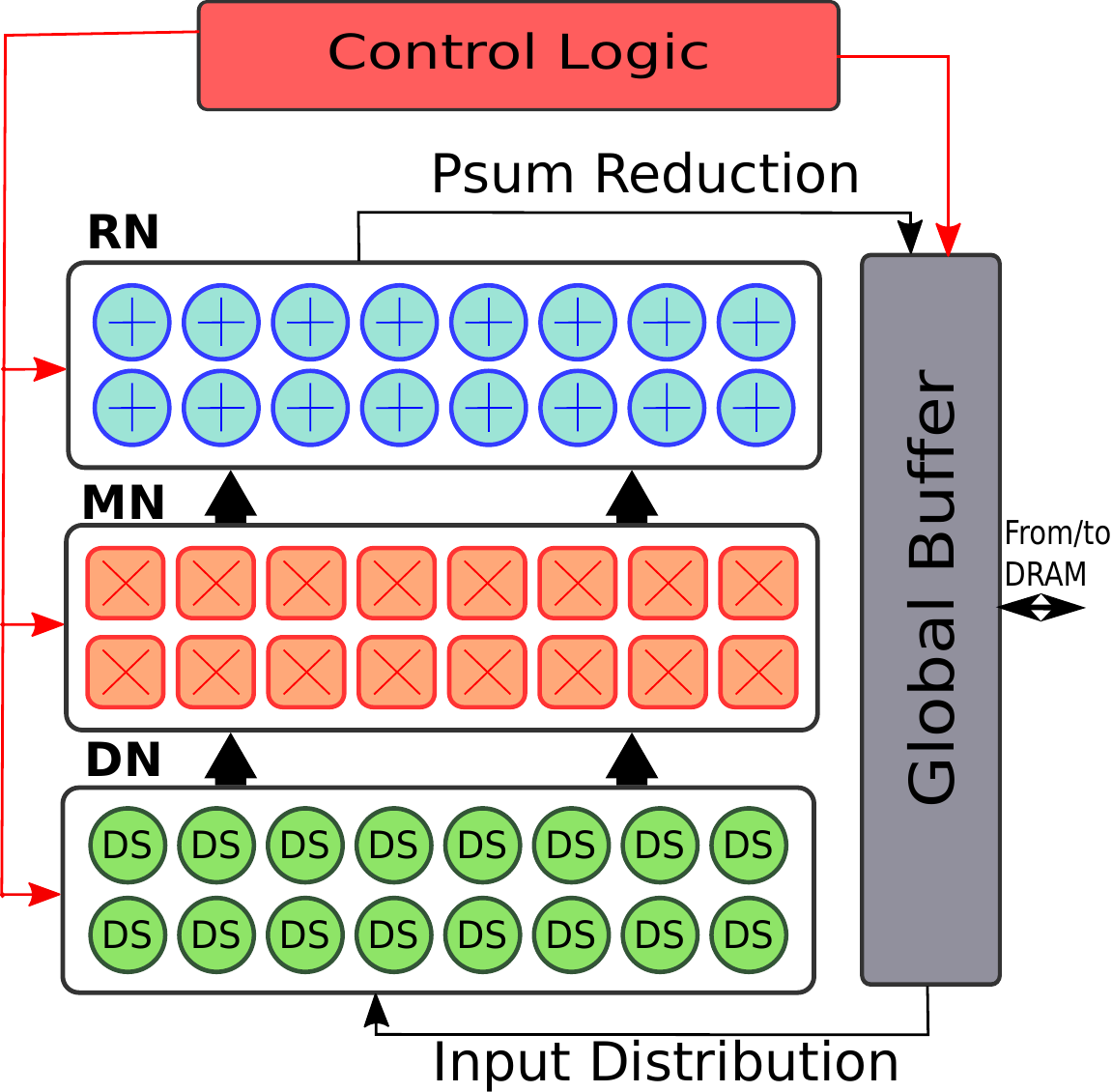}
        \end{center}
        \vspace{-0.2cm}
        \caption{Overview of the general flexible DNN inference accelerator architecture that is implemented as baseline in STONNE.}
        \label{fig:stonne_architecture}
\end{figure}

As previously commented, STONNE emerges as the first cycle-accurate simulation tool that enables exploration of the design space of flexible accelerator architectures. In this section, we explain the general flexible DNN inference accelerator architecture that is implemented as baseline in STONNE and whose high-level diagram is shown in Figure~\ref{fig:stonne_architecture}.


STONNE is equipped with all the major basic components of any recently proposed next-generation DNN accelerator~\cite{MAERI2018, SIGMA2020}:
i) A set of \textbf{Multipliers and adders} to carry out the multiply-and-accumulate operations required by each of the NN's neurons. ii) A \textbf{Memory hierarchy} composed of local storage, some on-chip/on-package global storage, and some connectivity to off-chip DRAM memory. Local storage is made up of buffers or registers that are typically private to certain groups of multipliers and adders, and are used to temporarily hold both input data (weights, activations and psums) and output results. On-chip/on-package global storage is typically shared by all multipliers and adders in the accelerator, and can be composed of either a single memory element (i.e., the Global Buffer, GB) or a hierarchy of different levels of memory (e.g., a cache hierarchy). These memory elements can be software- or hardware-managed. 
Off-chip DRAM memory can be private to the DNN accelerator or shared with a host compute platform such as a CPU. 
And finally, iii) \textbf{Control logic} that is used to reconfigure the connectivity among the compute and/or memory elements within the previous two components, thereby the architecture of the DNN inference accelerator can be made flexible and adaptable to efficiently map any compute/memory partitions and dataflows.

All the on-chip components are interconnected by using a three-tier network fabric composed of a Distribution Network (DN), a Multiplier Network (MN), and a Reduce Network (RN). These networks must accomplish certain requirements in order to provide the flexibility that the simulator promises. First, to compute all the MAC operations of a certain DNN layer, the DN distributes the required weights, activations or partial sums from the GB towards the MN. To enable all types of dataflows, the DN must provide support for unicast, multicast and broadcast data delivering. After the distribution, the multipliers at the MN carry out the multiplication operations generating the operands of the partial sums to be accumulated, and finally the RN network is equipped with adders that implement the required accumulations. Again, to enable dataflow flexibility, the RN must be capable of reducing any number of multiplier clusters of any size simultaneously.

\section{MAERI Architecture}
\label{section:architecture}
In this work, we have created an instance of the previous general flexible DNN inference accelerator architecture that corresponds to the MAERI architecture~\cite{MAERI2018}. This instance will be utilized for description and evaluation in the rest of the sections of this work. Note that, as our simulator is highly-modular, highly-configurable and highly-extensible by design, we can easily modify any of the above components to model any other type of accelerator architecture. 

\begin{figure*}[t!]
        \begin{center}
                \includegraphics[width=0.7\textwidth]{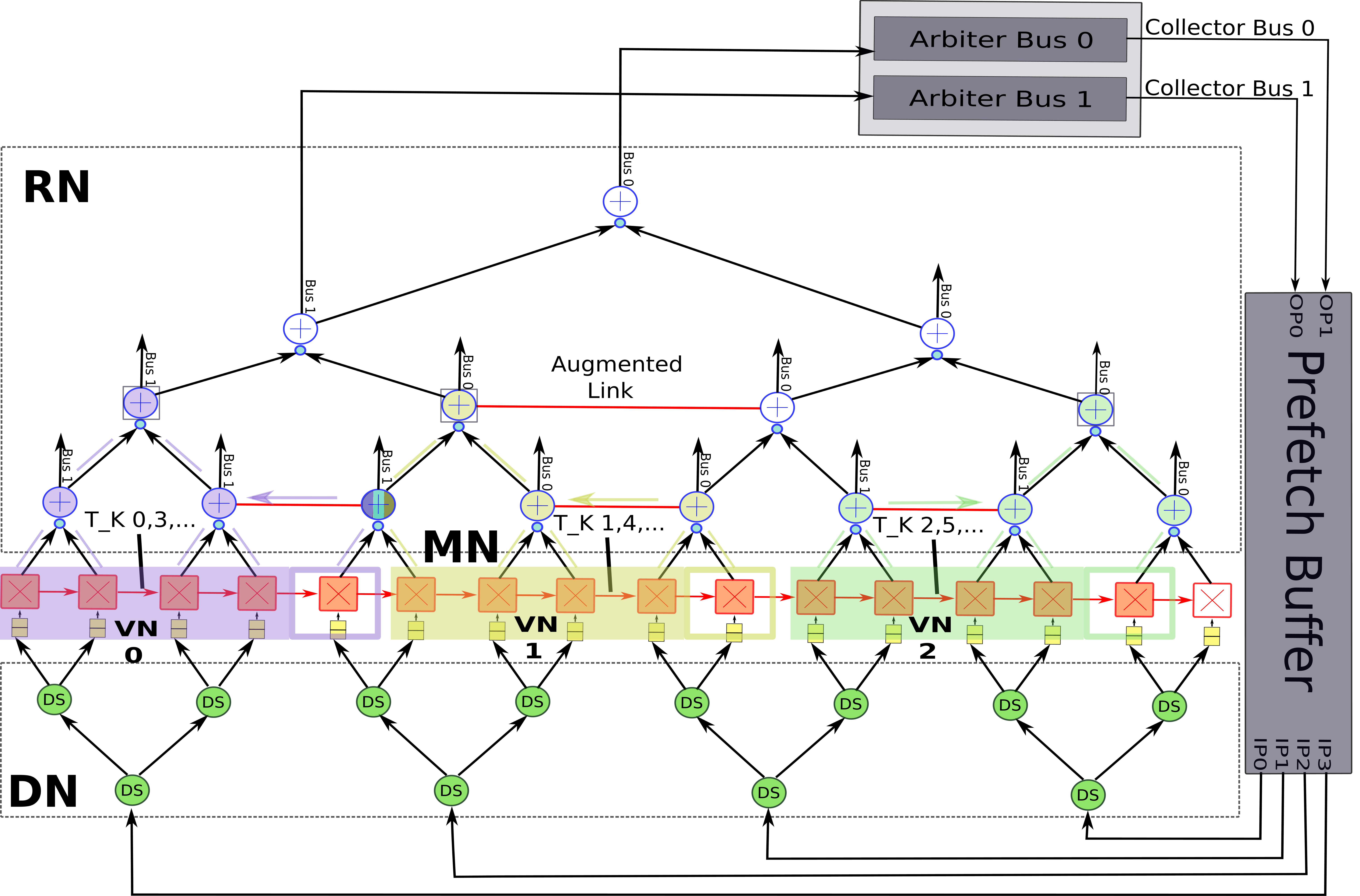}
        \end{center}
        \vspace{-0.2cm}
        \caption{Overview of the MAERI architecture. There are three Virtual Neurons (VN 0, 1 and 2) mapped on five different multipliers each. The fifth multiplier in each VN is the one that would be used to enable folding.}
        \label{fig:maeri_scheme}
\end{figure*}

An overview of the BSV-coded implementation of the MAERI architecture~\cite{MaeriCode} that we have faithfully modeled in STONNE is shown in Figure~\ref{fig:maeri_scheme}. 
At a high-level, there are two tree-based topologies that implement both the DN and RN networks, and 1D-mesh point-to-point network for the MN network. The Global Buffer is called Prefetch Buffer (PB) in MAERI. The PB needs arbitration at the write ports. The figure also illustrates an example of mapping three Virtual Neurons (VNs) on five different multipliers each. A VN is the most basic primitive in MAERI and is, in essence, a configurable cluster of multipliers used to execute the multiply operations in a certain output neuron. In principle, any VN could be set up using any number of multipliers.

\subsection{Flexible Network Fabric}
\label{subsec:maerinocs}
Next, we detail the implementation of the three networks (DN, MN and RN) in the MAERI architecture.

\textbf{Distribution Network (DN)}: As it is shown in Figure~\ref{fig:maeri_scheme}, the DN in MAERI is a binary-tree-based network topology that is replicated as many times as the number of read ports available in the PB. In the figure, the number of read ports is four so there are four sub-trees. Our simulator allows to configure the number of ports and sub-trees according to the user requirements. Each of these sub-trees is in charge of delivering the weights, inputs and psums from the PB to a different group of multipliers sited in the MN (explained below) through multicast, unicast, or broadcast operations. Each node of the DN is just a bufferless low-cost Distribution Switch (DS) that selects whether to send the input to one or both outputs using a bit vector that is set by the input source. 
Due to the simplicity of the DSs, the DN can provide single-cycle traversals from the PB to the MN for every piece of data.

\textbf{Multiplier Network (MN)}: This network is conformed by a set of Multiplier Switches (MSs) that can be configured to act as either forwarders or multipliers. The forwarding configuration is used to forward psums from the PB to the RN so that folding can be supported, whereas the multiplier configuration mode is utilized to compute a multiplication between a weight and an input value. In case folding is needed (further details in Section~\ref{sec:dnnmappings}), the architecture would need to allocate one extra MS for each VN to perform the forwarding of the psums calculated in the previous iterations of the same VN. 




\textbf{Reduce Network (RN)}: The RN is implemented as a tree-based topology augmented with one adder unit per node, a well-known layout for efficiently executing reduction operations. The tree structure is augmented with links between the nodes of the same level (horizontal links) that do not share the same parent. That is why the RN in MAERI is called \textit{Augmented Reduction Tree} (ART). More specifically, each node is a configurable Adder Switch (AS) that can be statically configured as either \textit{2:1 ADD}, \textit{3:1 ADD}, \textit{1:1 ADD plus 1:1 forward}, or \textit{2:2 forward}. This configurable capability within each RN node along with the augmented links are key aspects to enable high flexibility in MAERI, as they permit flexible support of multiple and non-blocking virtual reduction trees over a single physical tree hardware substrate.

As it is shown in Figure~\ref{fig:maeri_scheme}, each AS is connected to a different collector bus (there are two CBs in the figure). Each CB is used to write partial or final outputs in the PB. Note that, the number of CBs must be equal to the number of write ports in the PB. Since there are usually more ASs than collector buses, write requests often produce conflicts that will be managed by means of a bus arbiter module, which is shown at the top of the figure. 
Clearly, the higher the number of CBs and write ports the higher performance benefits that can be obtained due to lower network contention. However, this comes at the cost of higher energy consumption and on-chip area requirement.
In STONNE, we can easily study the impact of the number of output ports, CBs and their corresponding arbiters at design time according to what is needed. 

\subsection{\textit{Supported DNN mappings in MAERI}}
\label{sec:dnnmappings}
MAERI can be configured at execution time to run any number of VNs of arbitrary size. Basically, the DNN tile mapping taxonomy is similar to the one mentioned in~\cite{mRNA2019}. 
First, a layer is defined with 7 parameters as \textit{Layer(R, S, C, K, N, X', Y')} where R and S are the number of rows and columns in a filter respectively, C is the number of channels, K is the number of filters, N is the batch size, and X' and Y' are the number of rows and columns in the output respectively. Additionally, we have added in STONNE a new parameter (G) that allows MAERI to support factorized convolutions.


This way, we define a tile as \textit{Tile(T\_R, T\_S, T\_C, T\_G, T\_K, T\_N, T\_X', T\_Y')}, where ${T\_R}\times{T\_S}\times{T\_C}$ parameters are a subset of the filter dimensions, and therefore, what defines the size of the VN. Similarly, ${T\_G}\times{T\_K}\times{T\_N}\times{T\_X'}\times{T\_Y'}$ parameters represent the subset of number of groups, filters per group, input fmaps, and output dimensions, respectively, thus defining the number of VNs that are mapped onto the architecture. Note that, if the size of a VN cannot hold the entire filter size (i.e., $({T\_R}/{R}\times{T\_S}/{S}\times{T\_C}/{C})>1$), the architecture will have to enable what is called \textit{folding} as it will be necessary to iterate over the same VN, storing partial sums in some temporal storage (the prefetch buffer in the case of MAERI) and sending them back to the VN to be reduced with the calculated in the subsequent iteration. 

An example of tile mapping is depicted in Figure~\ref{fig:maeri_scheme}, where we have a Tile(T\_R=2, T\_S=2, T\_C=1, T\_G=1, T\_K=3, T\_N=1, T\_X'=1, T\_Y'=1) mapped into a MAERI instance and folding is enabled. Notice that with this tile shape, the number of mapped VNs is 3 (T\_K) and the VN size is 4 (${T\_R}\times{T\_S}$). As folding is needed, each VN would require one extra multiplier that would act as a forwarder (see the fifth multiplier allocated for each VN). 

\section{Validation and Evaluation}
\label{section:validation}

\subsection{Validation}
\label{subsec:validation}
To validate the MAERI architecture that we have simulated with STONNE against a real hardware implementation, we use the original BSV MAERI design~\cite{MaeriCode}.
For the validation process, we run STONNE using its direct user interface (the \texttt{STONNE User Interface} in Figure~\ref{fig:simulator_scheme}). Recall that this execution mode allows for easy configuration of the Simulation Engine (to model a MAERI architecture in this case), DNN layer configuration and memory/compute partition tiles.

Since the BSV MAERI version does not have the large flexibility of our cycle-accurate architectural simulator, which can model almost any combination of the parameters of the flexible accelerator (e.g., number of MSs, number of trees in DN, number of input/output ports in the Prefetch Buffer), we are heavily constrained in the number of validation experiments we can carry out. This way, we have configured both STONNE and BSV versions with 32 MSs and 4 DN/RN bw parameters. In other words, 4 input/output ports in the Prefetch Buffer, 4 trees in the DN (8 MSs per tree) and 4 Collector Buses.
In addition, the BSV version can only execute the three different types of layers listed in Table~\ref{table:layers_validation}, with the tile shape \textit{Tile(T\_R=3, T\_S=3, T\_C=1, T\_G=1, T\_K=1, T\_N=1, T\_X'=3, T\_Y'=1)}.

\begin{table}[t!]
\caption{Set of layers executed to validate the MAERI architecture simulated with STONNE.}
\label{table:layers_validation}
\begin{footnotesize}
\begin{center}
{
\begin{tabular}{|l|l|l|l|l|l|l|l|l|}\hline
\textbf{Name}                  & \textbf{R}   &   \textbf{S}   &   \textbf{C}   &  \textbf{G}  &   \textbf{K}   &   \textbf{N}   &   \textbf{X}   &   \textbf{Y}  \\\hline
TINY                  & 3   &   3   &   6   &  1  &   6   &   1   &   5   &   5  \\\hline
LATE\_SYNTHETIC       & 3   &   3   &   20  &  1  &   20   &   1  &    5   &   5  \\\hline
EARLY\_SYNTHETIC      & 3   &   3   &   6   &  1  &   6   &   1   &   20   &  20  \\\hline
\end{tabular}
}
\vspace{-0.65cm}
\end{center}
\end{footnotesize}
\end{table}

For functional validation of the STONNE-based MAERI simulator, we carry out an exhaustive head-to-head comparison between the STONNE and the BSV versions in terms of the output values produced by every single executed DNN layer in both platforms. After validation of all possible supported DNN layers with the above tile configuration, we can confirm that our  implementation of the MAERI architecture is correct.


\begin{figure}[t!]
        \begin{center}
                \includegraphics[width=0.40\textwidth]{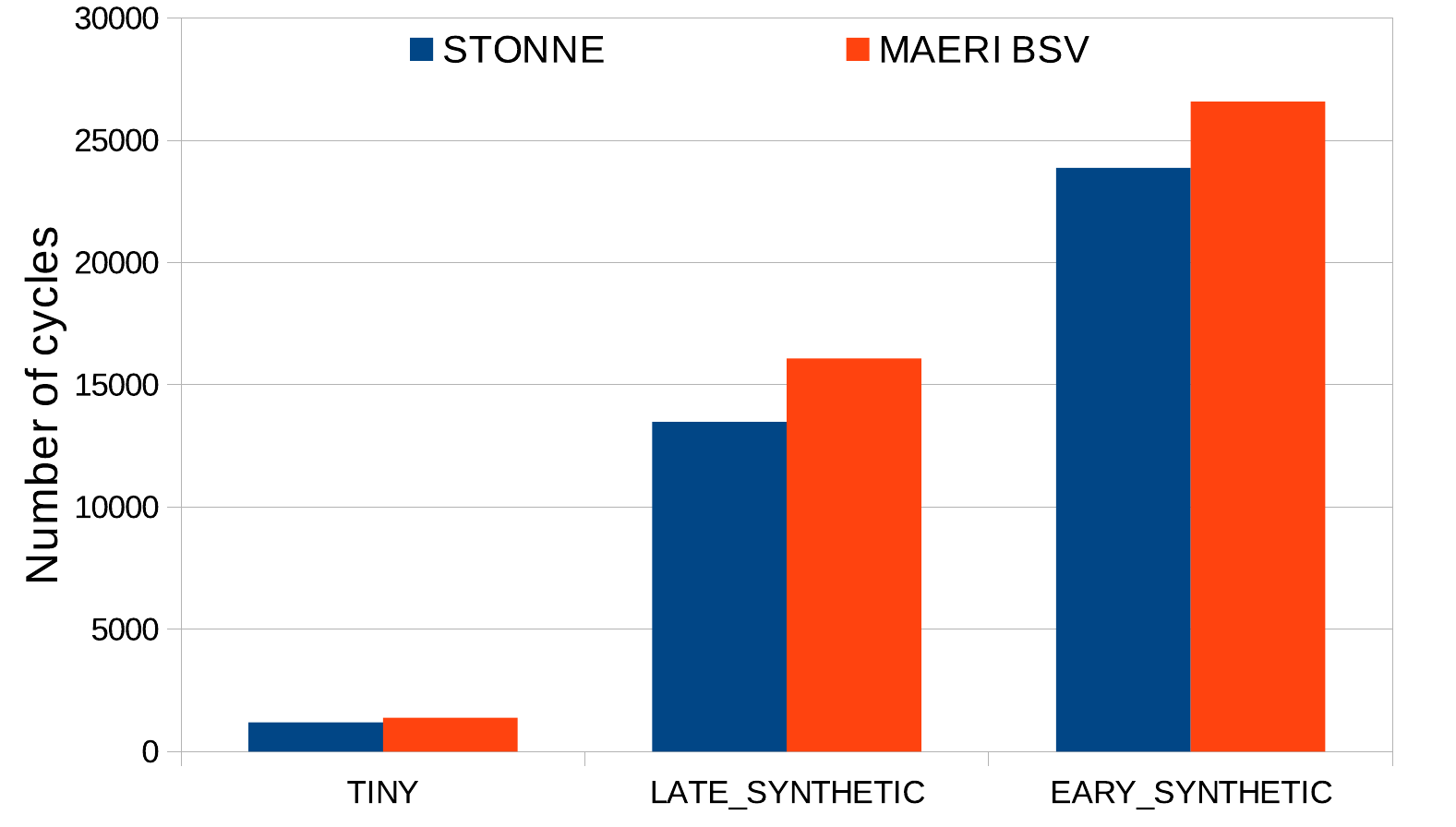}
        \end{center}
        \vspace{-0.2cm}
        \caption{Comparison in terms of number cycles between BSV MAERI and STONNE~for the layers described in Table~\ref{table:layers_validation}.}
        \label{fig:results}
\end{figure}

To evaluate the accuracy of timing simulation, Figure~\ref{fig:results} shows a comparison of the total number of executed cycles reported by the BSV MAERI and STONNE~versions after running the three types of layers supported by the BSV version (TINY, LATE\_SYNTHETIC and EARLY\_SYNTHETIC).
As we can see, our implementation of MAERI in STONNE shows an average difference in terms of total executed cycles of just 15\% as compared to the real BSV MAERI version (from  11\% to 19\%). These results demonstrate that STONNE closely mimic the characteristics of the hardware version. 
Additionally, after an in-depth analysis of the BSV MAERI code, we have found out that the small performance difference is mainly due to the testbench module used to run the BSV MAERI implementation. In particular, we discovered that the testbench does not fully leverage all the available input ports in parallel to feed the architecture.
That is the reason why our simulator achieves slightly better performance numbers (lower amount of clock cycles) for the two largest layers.

Finally, we have also validated the feasibility of the STONNE framework for conducting end-to-end evaluations  (see Table~\ref{table:state_of_the_art}). 
To do so, we have run five DNN models (AlexNet, MobileNets, Squeezenet, Resnet-50 and VGG-16) with a test set of 50 images from ImageNet, and for every image we have compared the score digits (output of the last fully-connected layer of each DNN) and predicted label reported by Caffe DL when running on a real back-end (CPU), with those obtained for the executions on the simulated MAERI architecture. We observed exactly the same results for all the images in both cases.

\subsection{Performance analysis of MAERI with real DNNs}
\label{subsec:perfanalysis}
The results previously shown demonstrate the correctness of the MAERI architecture being simulated by STONNE. Now, we show some of the key benefits of STONNE in providing detailed performance analysis when running real DNN models through its execution mode that enables end-to-end evaluation, i.e., the execution of real DNN models driven by a DL framework (Caffe in the current version of our simulation framework).
For the performance analysis, we simulate a baseline MAERI implementation with 64 MSs. On top of this simulated architecture we execute the five DNN models already mentioned (AlexNet, MobileNets, Squeezenet, Resnet-50 and VGG-16). For every layer, we obtain the optimal tile configuration reported by the mRNA tool and reconfigure the MAERI architecture accordingly before execution. Due to the large number of tile configurations needed for running all the layers in all the DNN models, we omit all the different tile configurations utilized for the sake of brevity. Note that, the BSV MAERI implementation cannot be used to carry out this design-space exploration as it only supports a particular tile configuration as commented in Section~\ref{subsec:validation}.

We have also performed simulations with 128 MSs but have not included the results since we have observed the same trends as for 64. 

\begin{figure}[t!]
        \begin{center}
                \includegraphics[width=0.5\textwidth]{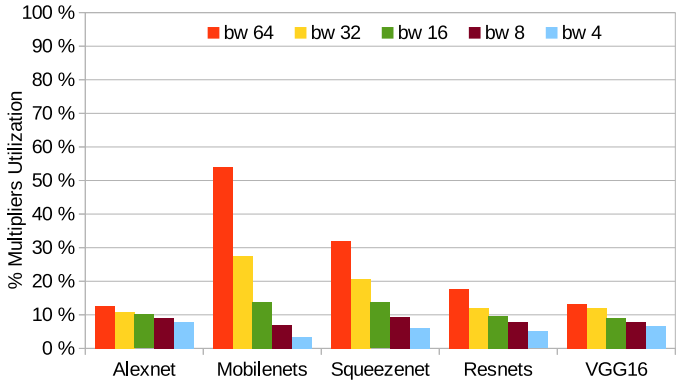}
        \end{center}
        \vspace{-0.2cm}
        \caption{Percentage of utilization in a MAERI-based architecture simulated using STONNE and using 64 multipliers.}
        \label{fig:utilization_64mses}
\end{figure}

One of the key aspects that reflects the efficiency in processing a particular DNN layer through a DNN inference accelerator is the resulting utilization of its compute resources, which depends on both the number of resources assigned to the configured tiles (theoretical utilization) and how these resources are actually leveraged. 

This way, guided by this observation, we first conduct a comprehensive analysis of the resources that are  mapped when processing the five DNN models in the simulated MAERI architecture with the baseline 64-MS configuration. Table~\ref{table:wasted_adders} shows the number of MSs that are mapped depending on the theoretical VN size (first column). Since our experiments reveal that folding is necessary for 98\% of the computation time, we assume that an extra MS is always needed to be able to map all the tiles (see the discussion in Section~\ref{sec:dnnmappings}) and therefore the real VN size is one unit larger (second column). We always map as many neurons as possible (third column).  The average frequency of every VN mapping configuration across the five DNN models is also shown in the fourth column. Note that, ordering the table data by frequency helps us understand the most common VN mapping configurations and the amount of idle resources left in each case.

As we can see in the fifth column of the table, we are far from reaching the near-100\% theoretical MS utilization that these flexible accelerator architectures for inference at the edge should approach. Specifically, the utilization of the MSs for the most frequent VN mapping configurations is very low. For example, a VN with a theoretical size of 36 MSs (the most typical VN configuration), and thus requiring a total of 37 MSs (the extra MS needed with this folding implementation), would leave 27 MSs unused, which results in a theoretical utilization of just 58\%.

\begin{table}[t!]
\caption{Theoretical MS utilization rate when mapping realistic VN sizes onto a 64-wide MAERI configuration.}
\vspace{-0.25cm}
\label{table:wasted_adders}
\begin{footnotesize}
\begin{center}
{
\begin{tabular}{|c|c|c|c|c|}\hline
\makecell{\textbf{Theoretical}\\\textbf{VN Size}} & \makecell{\textbf{Real}\\\textbf{VN Size}} & \textbf{\# of VNs} & \textbf{Frequency} & \makecell{\textbf{Theoretical}\\\textbf{MS Utilization}} \\\hline
\textbf{36} & 37 & 1 & 83\% & 58\%      \\\hline
\textbf{32} & 33 & 1 & 10\% & 52\%      \\\hline
\textbf{50} & 51 & 1 & 1\% &  78\%     \\\hline
\textbf{49} & 50 & 1 & 1\% &  76\%      \\\hline
\end{tabular}
\vspace{-0.25cm}
}
\end{center}
\end{footnotesize}
\end{table}
On the other hand, as it is explained above, this theoretical utilization is just the number of MSs that are mapped according to the tiles that have been used. However, the effective utilization depends not only on this, but also, on the capacity of leveraging all these mapped resources. 
Figure~\ref{fig:utilization_64mses} shows, for each one of the 5 DNN models considered in this work, the percentage of effective utilization for the 64 MSs as the input/output bandwidth is varied from 64 elements/cycle (maximum possible bandwidth) to 4 (see \textit{bw N} in the figure, where \textit{N} equals 64, 32, 18, 8 and 4). 

As we can see in Figure~\ref{fig:utilization_64mses}, MS utilization for all the five DNN models is particularly low in almost all cases: average utilization\footnote{Note that we count the multipliers that are being used as forwarders as effective utilization.} of 25\%, 16\%, 11\%, 8\% and 5\% for an input/output bandwidth of 64, 32, 16, 8 and \mbox{4 elements/cycle}, respectively.  Obviously, the lower the bandwidth at the PB's input and output interfaces, the higher the under-utilization of the 64 multipliers, which will be idle-waiting whereas contention arises at the DN and RN networks due to the limited number of input and output ports in the PB, respectively. However, even when considering the best hardware configuration (where the number of input and output ports equals the number of multipliers, i.e.  bandwidth of 64 elements/cycle), multiplier utilization results are also far from a near-60\%-utilization that the theoretical calculations previously discussed promised. In fact, utilization rate of multipliers is extremely low in both Alexnet and VGG-16, not even surpassing 10\%. 


\begin{table}[t!]
\caption{Configuration tiles used to run AlexNet DNN.}
\vspace{-0.25cm}
\label{table:alexnet_tiles}
\begin{scriptsize}
\begin{center}
{
\begin{tabular}{|l|l|l|l|l|l|l|l|l|}\hline
\textbf{Name}      &   \textbf{T\_R}   &   \textbf{T\_S}   &   \textbf{T\_C}   &   \textbf{T\_G}   &   \textbf{T\_K}   &   \textbf{T\_N}   &   \textbf{T\_X'}   &   \textbf{T\_Y'}  \\\hline
CONV1     &   11     &   11     &   1      &   1      &   1      &   1      &   1       &   1\\\hline
CONV2     &   5      &   5      &   2      &   1      &   2      &   1      &   1       &   1 \\\hline
CONV3     &   3      &   3      &   4      &   1      &   3      &   1      &   1       &   1 \\\hline
CONV4     &   3      &   3      &   6      &   1      &   2      &   1      &   1       &   1 \\\hline 
CONV5     &   3      &   3      &   6      &   1      &   2      &   1      &   1       &   1 \\\hline
FC6       &   1      &   12     &   1      &   1      &   8      &   1      &   1       &   1 \\\hline
FC7       &   1      &   16     &   1      &   1      &   4      &   1      &   1       &   1 \\\hline
FC8       &   1      &   8      &   1      &   1      &   10     &   1      &   1       &   1 \\\hline
\end{tabular}
\vspace{-0.35cm}
}
\end{center}
\end{scriptsize}
\end{table}

\begin{figure}[t!]
        \begin{center}
                \includegraphics[width=0.5\textwidth]{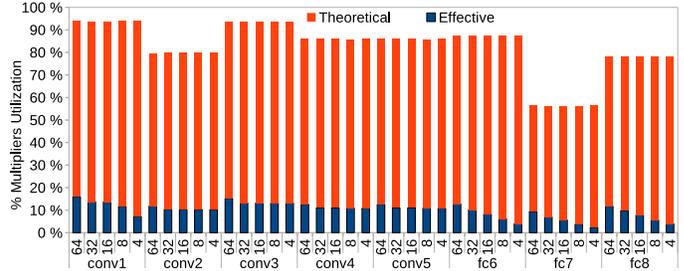}
        \end{center}
          \vspace{-0.2cm}
        \caption{Percentage of multipliers (MS) utilization of the baseline MAERI architecture simulated in STONNE with 64 MSs after execution of AlexNet. We compare the reported utilization (Effective) with the maximum achievable (Theoretical).}
        \label{fig:utilization_alexnet}
\end{figure}

We noticed that this performance issue becomes more evident when analyzing the percentage of multiplier utilization for single layers. Figure~\ref{fig:utilization_alexnet} shows these results for every layer in the case of AlexNet (the DNN model that reports the lowest utilization). 
Table~\ref{table:alexnet_tiles} shows the different tile configurations employed to run every one of the layers in AlexNet. In the figure, we overlap the effective multiplier utilization rate (see blue bars) with the maximum theoretical multiplier utilization that could be achievable according to the tile configurations. 
Clearly, multiplier utilization is extremely low compared with the theoretical value that could be achieved. Even in the configurations with no bandwidth restrictions (\texttt{bw 64}), the maximum utilization achieved is 15\% (CONV1 and CONV3) with some layers experimenting an utilization below 9\% (FC7).
This significant difference between the theoretical and effective utilization is, as explained in Section~\ref{section:architecture}, due to the dependency between the MN and the RN introduced by the psum when folding is used. This, impedes to iterate over the same output neuron in a pipeline manner,  hurting the utilization of the mapped resources and significantly degrading performance as was shown in Figure~\ref{fig:performance_speedup}.
This demonstrates the importance of properly supporting folding in a flexible accelerator architecture, as well as the need of a much more efficient implementation that allows for significant increase in the utilization rate of the compute resources.

\section{Conclusions}
\label{section:conclusions}

In this work we have presented STONNE, a cycle-accurate, highly-modular and  highly-extensible simulation framework that enables end-to-end evaluation of flexible accelerator architectures running complete contemporary DNN models. 
We have used STONNE to faithfully simulate a MAERI-based architecture with an average difference of only 15\% in total executed cycles. In addition, we demonstrate that the folding strategy implemented by the accelerator architecture is extremely inefficient, as it lowers compute unit utilization to an average of 25\% across 5 DNN models, which results into a maximum performance degradation of up to 610\%.

As part of our ongoing work, we are currently exploring the design of a novel reduction network capable of providing more efficient support to folding by avoiding the need to re-distribute the partial sums again once they reach the Prefetch Buffer. Additionally, we are extending STONNE to also report results of on-chip area and energy consumption based on the recently proposed Accelergy~\cite{accelergy2019} framework.




\section*{Acknowledgments} 

\footnotesize{The authors wish to thank Hyoukjun Kwon for for his clarifications on certain technical aspects related to MAERI. This work has been supported by the Spanish MCIU and AEI, as well as European Commission FEDER funds, under grant ``RTI2018-098156-B-C53''. Francisco Mu\~noz-Mart\'inez is supported by fellowship 20749/FPI/18 from Fundaci\'on S\'eneca, Agencia Regional de Ciencia y Tecnolog\'ia de la Regi\'on de Murcia.
}

\bibliographystyle{IEEEtran}
\bibliography{dnn-cal}



\end{document}